\documentclass[twocolumn,preprintnumbers,amsmath,amssymb]{revtex4}

\renewcommand{\d}{\textrm{d}}

\usepackage{graphicx}
\usepackage{dcolumn}
\usepackage{bm}

\usepackage{hyperref}
  \usepackage{latexsym}
  \usepackage{epsf}
  \usepackage{amssymb}

  \usepackage{amsmath}
  \usepackage{slashed,epsfig}

  \usepackage{bbm}
  \usepackage{bbm}
  \usepackage{amsmath,amssymb,amsthm}
  \usepackage{pst-all}
\newcommand{\be}{\begin{equation}}
\newcommand{\ee}{\end{equation}}
\renewcommand{\d}{\textrm{d}}

\def\gf{}

\begin{document}

\title{Euclidean axion wormholes have multiple negative modes}

\author{Thomas Hertog, Brecht Truijen, Thomas Van Riet }

\vspace{1.2cm}

\affiliation{{\small\slshape
Institute for Theoretical Physics, KU Leuven,\\
Celestijnenlaan 200D, B-3001 Leuven, Belgium   } \\
}

\begin{abstract}
We show that Euclidean axion wormholes in theories of gravity coupled to a single axion have several independent inhomogeneous perturbations that lower the Euclidean action. Our analysis relies on a judiciously chosen gauge-invariant variable which makes the negative mode structure about axion wormholes transparent. Perturbations lowering the action are concentrated in the neck region and exist for wormholes in flat space and in AdS. Their presence means axion wormholes are not relevant saddle points of the functional integral in quantum gravity. This resolves the paradoxes associated with these solutions from the viewpoint of AdS/CFT. 
\end{abstract}
\preprint{}

\maketitle

\section{Introduction}

Euclidean axion wormholes \cite{Giddings:1987cg,Lavrelashvili:1987jg} are regular extrema of the action in semiclassical quantum gravity theories that connect two distant regions or even two disconnected asymptotic regions. Despite much work over many years their status as saddle points of the functional integral remains unclear. Wormholes might be irrelevant, provide an imprint of other vacua of the landscape on our vacuum or give rise to an intrinsic randomness of the observed constants of nature (see e.g. \cite{Coleman:1988cy, Giddings:1988cx, Coleman:1989zu}) with interesting phenomenological implications for cosmology and particle physics (see \cite{Hebecker:2018ofv} for a recent review).

The Weak Gravity Conjecture \cite{ArkaniHamed:2006dz} lends some support to the physical relevance of wormholes because its generalization to instantons implies the existence of super-extremal instantons which, when sourced by axions, correspond to Euclidean axion wormholes \cite{Rudelius:2015xta, Brown:2015iha,Montero:2015ofa, Heidenreich:2015nta, Hebecker:2016dsw}. 

On the other hand, wormholes connecting two asymptotic regions are deeply puzzling in the context of AdS/CFT \cite{Maldacena:2004rf,ArkaniHamed:2007js}. From the field theory viewpoint the correlation functions across the two boundaries should factorize, while from the gravity point of view wormhole saddle point contributions would seem to introduce correlations between the two field theories.

One might have hoped that string theory would have avoided the paradoxes associated with Euclidean axion wormholes by not producing effective low energy theories allowing wormhole solutions. After all the existence of regular wormhole solutions depends delicately on the number of axions and dilatons in the theory and their couplings. However that does not appear to be the case. In recent years clean embeddings of axion wormholes in string theory have been given \cite{ArkaniHamed:2007js,Hertog:2017owm, Ruggeri:2017grz} further sharpening the paradox with AdS/CFT.

In this paper we provide evidence for a natural resolution of this paradox by analysing the behavior of the action in the neighbourhood of Euclidean axion wormholes (see \cite{Rubakov:1996cn, Rubakov:1996br, Kim:1997dm, Kim:2003js, Alonso:2017avz} for earlier, inconclusive work on this). We work with four-dimensional theories of gravity coupled to a single axion and a cosmological constant $\Lambda \leq 0$. In these theories we show that macroscopic axion wormholes which are everywhere in the semiclassical regime always have several independent inhomogeneous perturbations that lower the Euclidean action. This strongly suggests that axion wormholes are not relevant saddle points of the Euclidean functional integral.

To analyse whether there are physically meaningful perturbations that lower the action requires a careful choice of perturbation variable for which the Euclidean action is well-behaved and bounded below for normalized fluctuations.
The theories we consider have a single physical scalar perturbation degree of freedom that is a combination of the axion and scalar metric perturbations about these wormholes. We identify the conjugate momentum $\Pi_{\mathcal{X}}$ of the gravitationally dressed gauge-invariant axion perturbation $\mathcal{X}$ as a suitable variable which makes the negative mode structure about wormholes transparent. 

To derive the perturbation action we follow the functional integral procedure developed in a series of papers \cite{Gratton:1999ya,Gratton:2000fj,Gratton:2001gw} in which the degrees of freedom are identified in Lorentzian signature before the perturbation action is continued to the Euclidean. The boundary condition that the axion charge remains constant at the mouth of the wormhole implies vanishing Dirichlet boundary conditions on $\Pi_{\mathcal{X}}$. We find this sets the homogeneous perturbation to zero, which in turn means the conformal factor problem does not interfere with our analysis.

The action of perturbations that are inhomogeneous on three spheres at constant scale factor is manifestly positive in the asymptotic regions of the wormhole, as one expects from the field theory limit. The kinetic term is positive everywhere but the potential is negative (though bounded below) in the near neck region. It is this feature, which is particular to axions, that gives rise to multiple independent perturbations concentrated in the near neck region that lower the action. 

The presence of negative modes often points towards the existence of a lower action saddle point. The fact that the perturbations lowering the action are inhomogeneous, however, suggests that in this case macroscopic wormholes fragment ultimately breaking up the connection between the asymptotic regions.

\section{Axions in Euclidean space}\label{sec:axions}
 Axionic scalars $\chi$ have a shift symmetry at the level of the classical action. The axion charge $Q$ under this symmetry is related to the radial derivative of the axion,
 \be \label{Q}
 Q \sim g^{rr}\sqrt{g}\partial_r \chi\,.
\ee 
Axion Euclidean wormhole solutions are usually interpreted as instantons that violate axion charge conservation. This is because half wormholes describe the creation or absorption of baby universes, a process which from the point of view of the mother universe amounts to the loss or creation of axion charge. Wormholes in this interpretation must be viewed as saddle points of the path integral in momentum space since the boundary condition is that one fixes $Q$ at both wormhole ends \cite{Burgess:1989da} which, eq. \eqref{Q} shows, amounts to fixing the Euclidean axion momentum.
Schematically, the axion charge transition amplitude reads
\be \label{trans}
K \equiv \langle \Pi_F | \exp(-HT)|\Pi_I\rangle\,,
\ee
where $|\Pi \rangle$ are axion \emph{momentum} eigenstates, defined via a functional Fourier transform:
\be
|\Pi\rangle  = \int \d[\chi]\,e^{i\int_{\Sigma}\,\chi  \Pi } |\chi\rangle\ .
\ee
Here $\Pi $ is the Euclidean-time or `radial' component of a one-form that is orthogonal to the spacelike slice $\Sigma$. Hence $|\Pi\rangle=|Q\rangle$, and momentum eigenstates are charge eigenstates.

It is well known (see e.g. \cite{Burgess:1989da, Bergshoeff:2005zf, ArkaniHamed:2007js, Hebecker:2018ofv}) that axions have a kinetic term with the wrong sign in Euclidean signature. Equivalently, one can consider imaginary configurations in theories with normal sign kinetic terms as contributions to the transition amplitude \eqref{trans}. However it is arguably more appealing to regard the Hodge dual formulation of the axion as fundamental in which the axion enters as a $(D-1)$-form field strength $F$ and the action reads
\be
S = - \tfrac{1}{2\kappa^2} \int\sqrt{|g|}\Bigl(\mathcal{R} - \frac{1}{2} \frac{1}{(D-1)!}F^2  - \Lambda\Bigr)\,.
\ee
In this formulation, the axion has a standard kinetic term and the saddle points are real solutions.  
Hodge duality transforms this action into the following,
\be\label{action}
S=  -\tfrac{1}{2\kappa^2} \int\sqrt{|g|}\Bigl(\mathcal{R} + \frac{1}{2}\partial\chi\partial\chi  - \Lambda + \nabla(\chi\partial\chi ) \Bigr)\,. 
\ee
This action has a wrong sign kinetic term, at least when the axion is taken to be real. It also has an additional total derivative which provides an important contribution to the on-shell action.

\section{Wormhole solutions}
We consider the simplest possible model with axionic wormholes, consisting of gravity coupled to a single axion and a negative cosmological constant $\Lambda = -\frac{\alpha}{l^2}$ where $l^2$ is the AdS radius and $\alpha=(D-1)(D-2)$. 
We write the metric of spherically symmetric wormhole solutions as
\begin{equation} \label{wormholegeometry}
\d s^2 = f(r)^2\d r^2 + a(r)^2\d \Omega_{D-1}^2\,,
\end{equation}
together with the axion profile $\chi(r)$. The axion equation of motion $\Box \chi=0$ can be conveniently solved in terms of the radial harmonic function $h(r)$ defined as $\d h= f a^{1-D}\d r$. One finds that $\chi \propto h$ where the constant of proportionality squared, 
\begin{equation}\label{const}
\left(\frac{\d\chi}{\d h}\right)^2 \equiv  c>0\,,
\end{equation}
is precisely the axion charge $Q$ squared, i.e. the constant of motion associated with the shift symmetry of the axion. Using this the Einstein equation reduces to a single first-order equation for the scale factor \cite{Gutperle:2002km, Bergshoeff:2005zf},
\begin{equation}\label{firstorder}
\left(\frac{a'}{f}\right)^2= 1 + \frac{a^2}{l^2} -\frac{c}{2\alpha}a^{-2(D-2)}\,,
\end{equation}
where $a' \equiv \partial_r a$. Hence a wormhole solution can be written as
\begin{equation}\label{metric}
\d s^2 = \Bigl( 1+ \frac{\tau^2}{l^2} -
\frac{c}{2\alpha}\tau^{-2(D-2)}\Bigr)^{-1} \d \tau^2 + \tau^2\d \Omega^2\,.
\end{equation}
The solutions clearly asymptote to Euclidean AdS (EAdS) at large $\tau$ regardless of the value of $c$. When $c=0$ the solution is EAdS everywhere. In models \eqref{action} with a normal scalar rather than an axion, turning on the scalar yields $c<0$ which gives a singular solution that is a spike-like deformation of EAdS \cite{Bergshoeff:2005zf}. By contrast for axions $c>0$ and the solutions describe smooth wormholes with a minimum value $a(r=0)=a_0$ of the scale factor given by 
\be
a_0 = \left(\frac{c}{2\alpha}\right)^{1/(2D-4)} +{\cal O}(c/l^{2(D-2)})^2\,.
\ee
Macroscopic wormholes which are everywhere in the semiclassical regime must have $a_0 \gg 1$ in Planck units and hence require $c \gg 1$. The regularity of the solutions at the neck means one does not expect these wormholes to carry a net axion charge. One can rather think of them as charge conduits \cite{Heidenreich:2015nta}, where both mouths carry opposite charges. The regularised Euclidean action of a wormhole is given by
\be\label{action2}
S = \frac{\text{Vol}(S^{D-1})}{2\kappa^2}\sqrt{\tfrac{2(D-1)\pi^2}{(D-2)}}\,\sqrt{c} +\mathcal{O}(a_0/l) \,,
\ee
where $\sqrt{c}$ is proportional to the quantised axion charge $Q$.

In theories with multiple axions and saxions (dilatons) the axion kinetic term is replaced by a sigma-model, $G_{ij}\partial \phi^i\phi^j$, of indefinite signature. The instantons are then geodesic curves with $-c$ equal to the geodesic velocity squared \cite{Hertog:2017owm}.

\section{Negative modes}

We now turn to the behavior of the action around the axionic Euclidean wormhole saddle points. We focus on wormholes in four dimensions, but our results likely generalize to other dimensions $D>2$. As discussed above the action of quadratic perturbations of wormholes yields crucial input to elucidate the role -- if any -- of these wormholes in semiclassical quantum gravity. In particular the existence and number of negative modes around the wormholes should, we believe, be a major consideration in their interpretation \cite{Coleman:1987rm,Rubakov:1996cn,Gratton:2000fj}.

To determine whether there are physically meaningful negative modes it is imperative to work with well-behaved perturbation variables for which the Euclidean action is bounded below for normalized fluctuations. We follow the functional integral procedure developed in a series of papers \cite{Gratton:1999ya,Gratton:2000fj,Gratton:2001gw} in which the physical degrees of freedom are carefully identified in Lorentzian signature before the perturbation action is continued to the Euclidean.

\noindent {\it Perturbation action} \ We write the line element of a general scalar perturbation of the Lorentzian continuation of the wormholes as follows,
\begin{align}
\d s^2 = & b^2\Bigl( -(1+A)^2 \d \eta^2 + \partial_i B \d x^i \d\rho\,\, + \nonumber\\
&[(1-2\psi)\gamma_{ij} + \partial_i\partial_j E]\d x^i\d x^j\Bigr)\ ,
\end{align}
where we have adopted conformal gauge $f=a$, with $b$ the conformal scale factor and $\eta \in [0,\infty[$ conformal time, covering half the wormhole. The three-metric $\gamma_{ij}$ is the metric on $S^3$. Together with the fluctuation $\delta\chi$ of the scalar (axion) field this results in five scalar perturbation fields in total. 

Our starting point is the second order Lorentzian action for a general linear scalar perturbations of this kind \cite{Gratton:1999ya}. The constraints enforcing gauge invariance are imposed by introducing canonically conjugate momenta to rewrite the action in first-order form, and then functionally integrating over the non-dynamical fields. One then has some choice in deciding which linear combination of the remaining variables to use in order to describe the single physical scalar degree of freedom of the system. A suitable variable which makes the negative mode structure transparant is one whose Euclidean action is bounded below for normalized fluctuations around the backgrounds in question that obey the appropriate boundary conditions. This condition corresponds to having a positive kinetic term in the Euclidean action for all values of the Laplacian $\Delta$ on the three-sphere. It turns out that for axion wormholes a good physical choice of variable is the gauge-invariant combination $\mathcal{X}$ defined as 
\be
\mathcal{X} = \psi + \frac{b'}{b\chi'}\delta \chi\,,
\ee
where prime is now a derivative with respect to conformal time $\eta$. Substituting $\delta \chi$ for $\mathcal{X}$ in the action and performing the remaining functional integrals we get the action of $\mathcal{X}$ \cite{Gratton:2001gw}.

It is convenient to decompose the perturbations in modes $\mathcal{X}_n$ multiplying spherical harmonics on the three-sphere where $n$ labels the eigenvalues $-n^2+1$ of $\Delta$, with $n>0$. The complete set of fluctuation modes divides into the $n=1$ mode which is $O(4)$ invariant in the Euclidean region and the $n>2$ modes which describe inhomogeneous perturbations on the three-sphere. The $n=2$ mode is pure gauge. Continuing the action for $\mathcal{X}_n$ to the Euclidean and defining $\rho=i\eta$ yields \cite{Gratton:2001gw}
\be \label{quadr}
S_2 =\frac{\text{Vol}(S^3)}{\kappa^2}\int \d\rho \bigl(  A_n \dot{\mathcal{X}}_n^2 - B_n \mathcal{X}_n^2\Bigr)\,,
\ee
where a dot is a derivative with respect to Euclidean time $\rho$ and the functions $A_n$ and $B_n$ are given by
\begin{align}
& A_n = \frac{b^2 z^2}{1+z^2/(4-n^2)}\,,\\
& B_n = A_n \times [\frac{2}{1+z^2/(4-n^2)}\frac{\dot{z}b}{z\dot{b}}+ 5-n^2]\,,
\end{align}
with $z \equiv \frac{\dot{\chi} b}{\dot{b}}$ and $n \neq 2$. This variable $z$ differs by a factor $i$ from the corresponding variable in \cite{Gratton:2001gw} which considers saddle points involving a scalar instead of an axion. The overall sign of the action \eqref{quadr} follows from the requirement that perturbations must be well-behaved away from the neck region or, more generally, in the limit where gravity decouples \footnote{We thank Arthur Hebecker and Pablo Soler for discussions of this point.}.

We emphasized earlier that axionic wormholes should be viewed as saddle points of the Euclidean path integral in momentum space, with vanishing Dirichlet boundary conditions on the axion momentum. Our variable $\mathcal{X}$ is the gravitationally dressed axion. To complete the derivation of the perturbation action we therefore express the action \eqref{quadr} in terms of the conjugate momentum  $\Pi_{\mathcal{X}}^n$,  
\be\label{momact}
S_2 = \frac{\text{Vol}(S^3)}{\kappa^2}\int \d\rho\bigl( - B_n^{-1}(\dot{\Pi_{\mathcal{X}}^n})^2 + A_n^{-1}(\Pi_{\mathcal{X}}^n)^2\bigr)\,.
\ee 

\noindent {\it Stability analysis}\,. 
The kinetic term in this action is everywhere positive. The potential is bounded from below and negative only in the near neck region. Moreover the action has no divergences, and the gradient terms contribute positively. Hence the momentum variable $\Pi_{\mathcal{X}}^n$ is a good physical choice of variable in which the negative mode structure around axionic wormholes can be analysed.

In the absence of a cosmological constant the coefficients $A_n$ and $B_n$ in the action \eqref{momact} follow from eqs. \eqref{const} - \eqref{metric} that specify the wormhole backgrounds and read
\begin{align}
 A_n = &
\sqrt{\frac{|c|}{12}}\frac{(4-n^2)\cosh(2\rho)}{3 + (4-n^2)\sinh^2(2\rho)}\,,\\
 B_n = & \sqrt{\frac{|c|}{12}}\frac{(4-n^2)\cosh(2\rho)}{3 + (4-n^2)\sinh^2(2\rho)}\times \\\nonumber &[\frac{-4(n^2-4)\cosh^2(2\rho)}{(n^2-4)\sinh^2(2\rho)-3} - n^2 + 5 ]\,.
\label{ABwh}
\end{align}
For the homogeneous $n=1$ perturbation we have $B_1=0$ which implies $\dot\Pi_{\mathcal{X}}^1=0$ everywhere, consistent with earlier results \cite{Alonso:2017avz} but in contrast with the non-gauge invariant treatment in \cite{Rubakov:1996cn}. This means there is no homogeneous propagating degree of freedom. This is precisely what one expects since the boundary conditions we selected on physical grounds fix the charge $Q$, leaving us with the Hamiltonian constraint (\ref{firstorder}) only. This also means we do not encounter a conformal factor problem with these boundary conditions, which resonates with \cite{Witten:2018lgb}. 

\begin{figure}[h!]
\begin{center}
\includegraphics[width=0.3\textwidth]{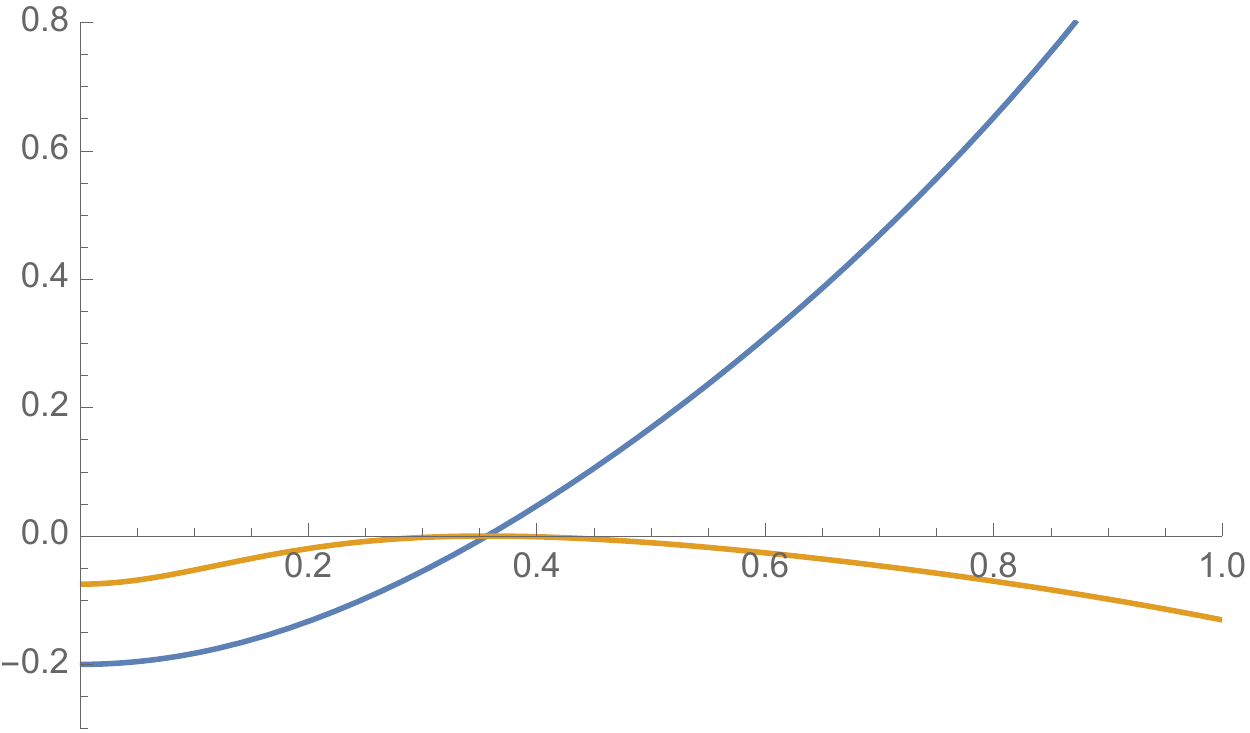}
\caption{\small \emph{The coefficients $A_n^{-1}$ (blue) and $B_n^{-1}$ (orange) entering in the action for perturbations about axion wormholes, shown here for $n=3$ (and with $c=1$).}} 
\end{center}
\end{figure}
Next we consider the action of inhomogeneous $n>2$ perturbations. Fig 1 shows the functions $A_n$ and $B_n$ for $n=3$. Their behavior across the entire wormhole is qualitatively similar for all $n>2$. There is a singular point at $\rho=\rho^*$ where $A_n^{-1}$ and $B_n^{-1}$ both vanish. For $\rho>\rho^*$, in the asymptotic region of the wormhole where gravity is not important, the action is manifestly positive. By contrast, the potential is negative for 
$\rho < \rho^*$, in the neck region, leading to the possibility that perturbations confined to the neck region lower the action. The critical value $\rho^*$ is the same for all $n$. The eigenvalue equation associated with the action \eqref{momact} is of Sturm-Liouville form, but the actual diagonalisation of this operator is impossible. However it is easy to see there are small fluctuations obeying our boundary conditions that lower the action. An example is $\Pi_{\mathcal{X}}= \cosh^{-1}(4\rho)$. Given this applies to all $n>2$ and given the large degeneracy of the spherical harmonics on $S^3$ this means the action of macroscopic axionic Euclidean wormholes has multiple negative modes in the inhomogeneous sector. This strongly suggests they do not provide a physically meaningful contribution to axion charge transition amplitudes like \eqref{trans}.

When $\Lambda < 0$ the metric of the wormholes is only known numerically in conformal gauge. A numerical analysis of the functions in the action \eqref{momact} for nonzero $\Lambda$ shows they exhibit a qualitatively similar behavior. Hence also axionic wormholes connecting two asymptotically $AdS$ regions have multiple negative modes. By contrast, a similar perturbation analysis applied to the class of instantons with a spike at the neck that are sourced by a dilaton, i.e. a scalar without shift symmetry which enters in the Euclidean theory with a normal kinetic term, shows these have no negative modes.

\section{Discussion}

We have calculated the second order variation of the action of axion and scalar metric perturbations about macroscopic Euclidean axion wormholes in theories of gravity coupled to a single axion in flat space and in AdS. We have shown there are multiple inhomogeneous small perturbations that lower the Euclidean action of these wormholes. This means that axion wormholes do not provide relevant saddle point contributions to the functional integral in semiclassical quantum gravity specifying axion charge transition amplitudes.

Our perturbation analysis is based on two crucial insights which resolve the ambiguities plaguing previous calculations and which make the negative mode structure about wormholes transparent. First, we have identified a gauge-invariant variable, the conjugate momentum of the gravitationally dressed axion perturbation, for which the Euclidean action is everywhere well-behaved and bounded below. Second, we have imposed Dirichlet boundary conditions on this to account for the fact that wormholes are saddle points of the functional integral in momentum space. It would be interesting to generalize our analysis to axion dilaton wormholes which in the Euclidean have kinetic terms of the form $-(\partial\phi)^2 + e^{b\phi}(\partial\chi)^2$, where $\phi$ is the dilaton and $b$ a coupling constant. The addition of a dilaton allows for extremal D-instanton solutions,  around which perturbations should have a manifestly positive quadratic action.

The presence of negative modes about a saddle point is usually a perturbative indication of the existence of 
a saddle point with lower action. In the case at hand the fact that we find negative modes in the inhomogeneous sector suggests that macroscopic wormholes fragment. The lower action configuration one flows to {\gf might} consist of an ensemble of microscopic `quantum wormholes' with unit charge. 
This is suggested by time-like T-duality which relates axion wormholes to super-extremal objects with $Q>M$, which one expects fragmentate into super-extremal particles of unit charge because the gravitational pull is weaker than electric repulsion.

But quantum wormholes have Planckian sized necks and are not valid (regular) semiclassical saddle points. The above reasoning thus suggests that smooth geometric wormhole connections between distant regions are  basically broken in gravitational theories with axions, thereby plausibly resolving the various paradoxes with AdS/CFT associated with these. The classically singular nature of microscopic quantum wormholes also means our stability analysis does not apply to these. Hence our results leave open the possibility that unit charge wormholes enter as non-perturbative contributions to the axion potential, breaking the shift symmetry, with appealing implications for phenomenology.
 
\section*{Acknowledgments}
We thank R.~Alonso, B.~Cottrell, A.~Hebecker, K.~M.~-Lee, M.~Kramer, M.~Montero,  M.~Trigiante, P.~Soler and A.~Urbano for useful discussions and  E.~van der Woerd for initial collaboration. We also thank A.~Hebecker, P.~Soler and E.~Kiritsis for remarks on an earlier draft.  The work of TVR is supported by the FWO Odysseus grant G.0.E52.14N. The work of TH is supported in part by the National Science Foundation of Belgium (FWO) grant G092617N, the C16/16/005 grant of the KULeuven and the European Research Council grant no. ERC-2013-CoG 616732 HoloQosmos. BT is funded by an FWO PhD fellowship. We also acknowledge support from the COST Action MP1210 `The String Theory Universe'. 

\renewcommand{\tt}{\normalfont\ttfamily}
\bibliography{Wormholes}
\bibliographystyle{utphysmodb}
\end{document}